\documentclass[twocolumn,amsmath,amssymb,prl,10pt,nofootinbib,superscriptaddress]{revtex4}

\usepackage{caption}
\usepackage{subcaption}
\usepackage{graphicx}
\usepackage{dcolumn}
\usepackage{bm}
\usepackage{romannum}
\usepackage{amsmath}
\usepackage{amssymb}
\usepackage{float}
\usepackage[export]{adjustbox}
\usepackage[labelsep=period]{caption}
\usepackage[normalem]{ulem}
\usepackage{mathtools}
\usepackage{siunitx}
\usepackage{soul}
\usepackage{physics}
\usepackage{minitoc}

\usepackage{bm}
\usepackage{xcolor}
\usepackage{lipsum}

\begin{document}

\pagenumbering{arabic}

\preprint{APS/123-QED}

\title{ The gravito-optic effect}

\author{Eduard Atonga}%
\email{eduard.atonga@univ.ox.ac.uk} 
\affiliation{%
Department of Physics, Atomic and Laser Physics sub-Department, Clarendon Laboratory, University of Oxford,\\
Parks Road, Oxford OX1 3PU, United Kingdom
}%

\author{Ramy Aboushelbaya}%
\email{ramy.aboushelbaya@physics.ox.ac.uk}
\affiliation{%
Department of Physics, Atomic and Laser Physics sub-Department, Clarendon Laboratory, University of Oxford,\\
Parks Road, Oxford OX1 3PU, United Kingdom
}%

\author{Peter A. Norreys}%
\email{peter.norreys@physics.ox.ac.uk}
\affiliation{%
Department of Physics, Atomic and Laser Physics sub-Department, Clarendon Laboratory, University of Oxford,\\
Parks Road, Oxford OX1 3PU, United Kingdom
}%
\affiliation{%
John Adams Institute for Accelerator Science, Denys Wilkinson Building, University of Oxford\\
Keble Road, Oxford OX1 3RH, United Kingdom
}%

\begin{abstract}

Gravitational waves have predominantly been detected using interferometric techniques, with
standard approaches limited to 10 kHz and with modern advancements extending this bound to 300 kHz.
To explore the largely uncharted higher-frequency gravitational wave spectrum, a general wave
optics formalism is presented here, revealing a space-time-periodic correction to the electromagnetic
wave equation, along with the emergence of sidebands at characteristic angles in the presence of
an incident gravitational wave (analogous to acousto-optic diffraction). Then a
Fabry-Pérot-enhanced heterodyne detection scheme is proposed to amplify and measure this effect. It is demonstrated that such a detection strategy has the potential of extending gravitational wave detection to high-frequency regimes far beyond the LIGO sensitivity
band. This novel approach provides a pathway to probing new physics through high-frequency
gravitational waves in the MHz-GHz range.

\end{abstract}

\maketitle

\section{\label{sec:Introduction}Introduction}

The direct detection of gravitational waves has opened a new observational window into astrophysics and fundamental physics. While interferometric detectors such as LIGO and Virgo have probed gravitational waves in the Hz–kHz regime \cite{PhysRevLett.116.061102},  upcoming observatories, such as LISA \cite{PhysRevD.73.064030,Karsten_Danzmann_1996,Karsten_Danzmann_2003}
and Cosmic Explorer, \cite{PhysRevD.103.122004,hall2022cosmic,evans2021horizonstudycosmicexplorer} aim to explore gravitational waves in the frequency regime below the aforementioned gravitational wave observatories, leaving much of the spectrum beyond 10 kHz relatively unexplored.
Theoretical predictions suggest that high-frequency gravitational waves, exceeding 10 kHz to the optical frequency regime and beyond, could arise from early universe phenomena, exotic compact objects, or beyond-standard-model physics. Detecting such signals remains a significant challenge, as conventional interferometric techniques lose sensitivity at these frequencies.

A variety of methods have been explored to detect high frequency gravitational waves such as polarization rotation detectors \cite{
park2021observation,cruise2000electromagnetic}, Bulk Acoustic Wave devices (BAW) \cite{PhysRevLett.127.071102, BAW1} and  graviton-magnon resonance effect \cite{ito2020probing}. Also, particular interest has been devoted to detection schemes utilizing the inverse Gertsenshtein effect \cite{gertsenshtein1962wave}. This phenomenon involves the resonant conversion of gravitational waves into electromagnetic radiation mediated by an electromagnetic background or probe field. In principle, it is applicable across all frequencies. In the GHz frequency regime, the excitation of microwave resonator modes and heterodyne measurement techniques has been explored. For THz frequencies and above, planetary magnetospheres \cite{PhysRevLett.132.131402}, static magnetic fields, and laser-based fields have been proposed as conversion mediums, with photons subsequently detected using single-photon detectors. A comprehensive review of these detection methods and others can be found in \cite{Aggarwal:2020olq}. Despite these significant efforts, gravitational wave detection within these frequency bands remains elusive, as the existing schemes either fall short of the required sensitivity or have failed to identify potential signals within the targeted frequency range. However, there exists a promising approach for detecting gravitational wave signals beyond the 10 kHz frequency limit that involves levitated sensor detectors based on optically levitated multilayered dielectric microstructures. These interferometric detectors are being designed to search for gravitational waves from coalescing primordial black holes \cite{Raidal:2017mfl,Franciolini:2022htd,Ireland:2023avg,Dong:2015yjs} and grand-unified-theory-scale QCD axions produced through superradiance around black holes \cite{Brito:2014wla,Arvanitaki:2014wva,Yoshino:2013ofa,Sun:2020gem,PhysRevD.81.123530,PhysRevD.83.044026,PhysRevD.95.043001}. However, their operational frequency band is limited to the range of 10–300 kHz \cite{PhysRevLett.110.071105,PhysRevLett.128.111101}.

In this article, a novel interaction between gravitational and electromagnetic waves, demonstrating that incident plane gravitational waves induce light diffraction, is presented in full. Moving beyond the traditionally used ray optics approach which yields the Eikonal or transport equations used to describe the operations of interferometric and polarization rotation-based gravitational wave detectors, the wave optics formalism reveals a space-time-periodic correction to the electromagnetic wave equation from the gravitational wave background. The spacetime-periodic correction to the wave equation due to the gravitational wave background, gives rise to an analogous effect to acousto-optic diffraction \cite{Ghatak_Thyagarajan_1989} which generates optical sidebands at distinct angles.  A Fabry-Pérot cavity-enhanced heterodyne scheme is proposed to amplify this effect, offering a viable near-term path for high-frequency gravitational wave detection.  Complementary to existing techniques, this method extends observational capabilities into unexplored frequency regimes, enabling new searches for high-frequency gravitational waves and potential signatures of new physics.

\section{Theory}

To a large extent, the present-day knowledge of the universe relies on inferences drawn from observations of the propagation of electromagnetic waves. From radio waves to gamma rays, these observations have provided critical insights into the structure, composition, and evolution of the cosmos. Electromagnetic radiation has been essential in testing the predictions of the general theory of relativity by providing a means to observe how gravitational fields influence the path and properties of light. For example, gravitational lensing, where light from distant stars is bent by massive objects and even the detection of gravitational waves using Michelson interferometers to measure the subtle distortions caused to spacetime by gravitational waves. 

The general relativistic treatment of the propagation of electromagnetic radiation is often developed through the variational principle \cite{Hobson, Maggiore:2007ulw}. Here, in the absence of external current densities, the free-field Lagrangian density governing the dynamics of electromagnetic fields within a curved spacetime (characterized by a symmetric metric $g_{\mu \nu}$ with a metric signature of $(1,-1,-1,-1)$) is expressed as a subset of the total Einstein Maxwell Lagrangian:

\begin{equation}
    \mathcal{L} \supset \frac{-1}{4\mu_0} g^{\mu \rho}g^{\nu \sigma}F_{\mu \nu}F_{\rho \sigma} 
\end{equation}

where, denoting $A^{\mu}$ as the vector potential , $F^{\mu \nu} = \nabla_{\mu} A_{\nu} - \nabla_{\nu}A_{\mu}$ is the anti-symmetric gauge invariant electromagnetic field tensor, which obeys the Bianchi identity $\nabla_{[\mu}F_{\rho \sigma]} = 0$ and  $\nabla_{\mu}$ is defined as the co-variant derivative. The associated Euler-Lagrange equations are thus compactly written as the co-variant divergence of the electromagnetic field tensor: 

\begin{equation}
    \nabla_{\mu} F^{\mu \nu} = 0. 
    \label{divergence}
\end{equation}

Taking the covariant derivative of this equation along with the Bianchi identity and considering the commutator of two co-variant derivative acting of a rank-2 tensor, assuming a torsion-less space-time, it follows that:

\begin{equation}
    [\nabla_{\alpha},\nabla_{\mu}] F^{\mu \nu} = R_{\lambda \alpha}F^{\lambda \nu} + R^{\nu}_{\lambda \alpha \mu} F^{\mu \lambda},
    \label{divergence}
\end{equation}
where $R_{ab} = g^{cd}R_{acbd}$ and $R_{abcd}$ are the Ricci and Riemann tensors, respectively \cite{Hobson,Carroll}. From this, it is now possible to derive the curved space-time wave equation of the electromagnetic field tensor:

    \begin{equation}
   \nabla_{\sigma} \nabla^{\sigma} F_{\mu \nu} + 2R_{\mu \sigma \nu \rho} F^{\sigma \rho} + R_{\mu}^{\sigma}F_{\nu \sigma} - R_{\nu}^{\rho}F_{\mu \rho} = 0. \label{BASE}
\end{equation}

 In most astrophysical applications, direct solutions of the electromagnetic field equations are rarely undertaken. Instead, for ease of computation, the ray optics approximation is often used, wherein the evolution of the wave vector, or the polarization vector, is computed to describe the propagation of electromagnetic waves through spacetime. Traditionally, this involves employing the Lorenz gauge ($\nabla_{\mu}A^{\mu}=0$) along with the introduction of an anazts for the vector potential \cite{bunney2021electromagnetism, park2021observation}.  
 
 Intriguingly, it has recently been shown that this analysis can be performed using the field tensor for which any ambiguity regarding the gauge invariance of the derived result is eliminated \cite{tsagas2004electromagnetic}. As a result, the following anazts $F_{\mu \nu} = f_{\mu \nu}(\bold{r})e^{i \Psi(\bold{r})}$ is introduced (where $\Psi(\bold{r})$ is a rapidly varying phase and $f_{\mu \nu}(\bold{r})$ is the spatially varying polarization amplitude).  Inserting the anazts into Equation (4) yields a solution that can be expanded in powers of the electromagnetic field wave vector $k_{\sigma} = \nabla_{\sigma} \Psi(\bold{r})$:

\begin{equation}
    k_{\mu} k^{\mu}f_{\sigma\rho} - 2i k^{\mu} \nabla_{\mu}f_{\sigma\rho} -i (\nabla^{\mu}k_{\mu}) f_{\sigma \rho} + \mathcal{O}(|k|^{0}) = 0 .
    \label{field}
\end{equation}

Employing the geometric optics approximation, where the wavelength of the electromagnetic radiation is much smaller than the characteristic length and time scale over which the curvature of spacetime varies $\lambda \ll g/\partial g$ and using the well-known slowly varying amplitude approximation \cite{powers2017fundamentals, new2011introduction} (where $|\nabla_{\sigma} f_{\mu \nu}(\bold{r})| \ll |k_{\sigma}f_{\mu \nu}(\bold{r})|$), to highest order in the electromagnetic wave number recovers the Eikonal and electromagnetic transport equation. 

Before reviewing the ray optics formalism in curved space-time, a brief discussion of the spatial profile of a laser beam profile is required. For a Guassian beam with a wavenumber $k_0$ and a sufficiently large beam width $w_0 $ (such that $ \epsilon = 1/k_0w_0 << 1$) the derivatives of the radial profile are negligible compared to the phase gradient. In addition to this, over the Rayleigh range $Z_R = 1/2 k_0 w_0^2$, the beam width is essentially constant. As a result, the spatial profile of a Gaussian beam over the Rayleigh range behaves as a plane wave with an effective beam radius $w_0$. It is then possible to further decompose the polarization amplitudes into a magnitude and polarization unit bi-vector $f_{\mu \nu}(\bold{r}) = \mathcal{A(\bold{r})}\zeta_{\mu \nu}(\bold{r})$. 

To order $\mathcal{O}(|k|^{2})$, the electromagnetic equations of motion yield the Eikonal equation $g_{\mu \nu}k^{\mu}k^{\nu} = 0$ and thus the phase gradient is null. Taking the covariant derivative of the Eikonal equation yields the null geodesic equations $\nabla_{\nu}(k_{\mu}k^{\mu} ) = k^{\mu} \nabla_{\mu} k_{\nu} = 0$. 
 To order $\mathcal{O}(|k|^{1})$ we get the transport equation $ k^{\mu} \nabla_{\mu}\zeta_{\sigma\rho}(\bold{r}) = 0$, which provides the evolution of the polymerization of the electromagnetic field along a geodesic. 

The application of the geodesic equation is well-established as the foundation for gravitational wave detection using Michelson interferometers, such as those employed on LIGO. The use of the transport equation in the context of gravitational wave detection has also been investigated, notably in the work of Adrian Cruise \textit{et al.} \cite{cruise2000electromagnetic}. However, detectors are not embedded in flat space-time but rather in a Schwarzschild spacetime due to Earth's curvature. For a planet of mass \( m \) and radius \( R \), since \( Gm / c^2 R \gg \mathcal{O}(10^{-21}) \) for Earth, distinguishing the background-induced polarization rotation \cite{Hobson,Carroll} from the cumulative polarization rotation caused by a passing gravitational wave would be challenging. The equations typically employed in gravitational wave detection reduce to the description of light to rays, which do not account for wave effects such as diffraction. To preserve a wave-based description of the electromagnetic field, instead of applying the ansatz \( F_{\mu \nu} = f_{\mu \nu}(\mathbf{r}) e^{i \Psi(\mathbf{r})} \) to Eq. (\ref{BASE}) as done previously, the covariant derivative is explicitly expanded in terms of partial derivatives and Christoffel symbols. Given that sources of gravitational waves are located at astronomical distances from the detector, their effects upon reaching Earth are notably weak \cite{Aggarwal:2020olq}. In addition, laboratory sources lack the energy densities required to induce significant deformations in the surrounding spacetime \cite{10.1063/1.4962520,Kadlecov,PhysRevD.110.044023}. This condition enables the expression of the metric tensor as the sum of the Minkowski metric and a metric perturbation denoted as $h_{\mu \nu}$, with the stipulation that $|h_{\mu \nu}| \ll 1$ for which:

\begin{equation}
        g_{\mu \nu} = \eta_{\mu \nu} +  h_{\mu \nu} \\.
\end{equation}

The contra-variant metric coefficients is obtained by imposing that $g_{\mu \sigma} g^{\sigma \nu} = \delta_{\mu}^{\nu}$. To first order in metric perturbation, this quantity is written as $ g^{\mu \nu} = \eta^{\mu \nu} -  h^{\mu \nu} + \mathcal{O}(h^2)$. Thus we are able to re-write eq.(\ref{BASE}) as:

\begin{widetext}
\centering
    \begin{eqnarray}
           \eta^{\beta \lambda}\partial_{\beta}\partial_{\lambda}F_{\tau \sigma} -h^{\beta \lambda}\partial_{\beta}\partial_{\lambda}F_{\tau \sigma} \nonumber \\-(\Gamma^d_{\tau \beta}\partial^{\beta}F_{d \sigma} +  \Gamma^d_{\sigma \beta}\partial^{\beta}F_{\tau d}) -\Gamma^{q \, \beta}_{\beta} \partial_{q} F_{\tau \sigma} -(\Gamma^{q}_{\tau \beta}\partial^{\beta}F_{q \sigma}+\Gamma^{q}_{\sigma \beta}\partial^{\beta} F_{\tau q}) \nonumber \\ 
   -(F_{d \sigma}\partial^{\beta}\Gamma^d_{\tau \beta} +  F_{\tau d}\partial^{\beta}\Gamma^d_{\sigma \beta})+ R_{\lambda \tau }F^{\lambda}_{\sigma}-R_{\lambda \sigma}F^{\lambda}_{\tau}  + 2 R_{\sigma \lambda \tau \mu}F^{\mu \lambda} = 0 
    \label{geometric optics master equation}
    \end{eqnarray}
\end{widetext}

There exist distinct regimes in which specific terms in Eq. (\ref{geometric optics master equation}) can be neglected. 

First, consider the interaction between an electromagnetic plane wave and a gravitational plane wave, characterized by frequencies $\omega_0$ and $\Omega$
respectively. In this context, the terms  $R_{\lambda \tau }F^{\lambda}_{\sigma}-R_{\lambda \sigma}F^{\lambda}_{\tau}$ vanish, given that gravitational plane waves are free-space solutions of the Einstein field equations in the absence of stress-energy sources, thereby satisfying  $R_{\mu \nu} = 0$. Consequently, for the intermediate frequency regime $\omega_0 \sim \mathcal{O}(\Omega)$ all remaining terms in Eq. (\ref{geometric optics master equation}) must be retained for a comprehensive analysis.

Second, we denote the high frequency regime as $\omega_0 \ll \Omega$; an example of this is the Gertsenshtein effect \cite{gertsenshtein1962wave}, in which gravitational waves are converted into electromagnetic waves via a static magnetic field ($\omega_0 =0$). In this regime, Eq. (\ref{geometric optics master equation}) reduces to:

\begin{eqnarray}
    \eta^{\beta \lambda}\partial_{\beta}\partial_{\lambda}F_{\tau \sigma}  = (F_{d \sigma}\partial^{\beta}\Gamma^d_{\tau \beta} +  F_{\tau d}\partial^{\beta}\Gamma^d_{\sigma \beta}) -  2 R_{\sigma \lambda \tau \mu}F^{\mu \lambda} 
\end{eqnarray}

The final regime is the converse $\omega_0 \gg \Omega$ and describes the low frequency regime. In this case Eq.(\ref{geometric optics master equation}) reduces to the simple expression:

\begin{equation}
    g^{\sigma \rho}\partial_{\sigma} \partial_{\rho} F_{\mu \nu} = 0
\end{equation}

Gravitational wave signals are typically not expected above a few kHz, with the upper bound set by the ring-down phase of neutron star mergers \cite{Aggarwal:2020olq, Bauswein:2015vxa}. However, high-frequency gravitational waves can also arise from various beyond-standard-model sources, including mergers or evaporation of light primordial black holes \cite{Raidal:2017mfl, Franciolini:2022htd, Dong:2015yjs, Ireland:2023avg} and superradiance phenomena \cite{Yoshino:2013ofa, Arvanitaki:2014wva, Brito:2014wla, Sun:2020gem}. A comprehensive review of gravitational wave sources and detection schemes in the MHz–GHz range is provided in \cite{Aggarwal:2020olq}. With the exception of primordial black holes, most sources do not exceed the \(10^{12}–10^{13}\) Hz range, while laboratory-generated gravitational waves from the most intense pulsed and continuous laser systems operate in the optical to UV regime (\(\omega_0 > 10^{14} \text{ Hz}\)) \cite{PhysRevD.110.044023}. The majority of beyond-standard-model sources are thus concentrated at GHz frequencies and below, thus the low frequency limit is the most relevant regime and will be the focus of this study.

The equation of motion for the electric field in linearized space-time in the low frequency limit can be derived by considering the observation of the electromagnetic field tensor by a stationary observer with four velocity $U^{\mu} = (c,0,0,0)$:

\begin{equation}
    (\Box - h^{\mu \nu}\partial_{\mu}\partial_{\nu})\bold{E}(t,\bold{r}) = 0 \label{DiffractionEOM}
\end{equation}

Where $U^{\mu} F_{\mu j} = E_j $ and $\Box = \eta^{\mu \nu}\partial_{\mu}\partial_{\nu}$ is the flat spacetime D'Alembertian differential operator. The dispersion relation for the electric field can also be readily derived by using the ansatz $\bold{E}(t,\bold{r}) = \bold{E}_0 \text{exp}(iK_0^{\mu}X_{\mu})$, where $K_0^{\mu}= (\omega_0/c, \bold{k}_0)$ and $X^{\mu}$ are the wave  and position four vector respectively. For this, one gets the dispersion relation in the  low frequency limit $\mathcal{D}(\omega_0, \bold{k}_0) =  g_{\mu \nu}K_0^{\mu}K_0^{\nu} = 0$.

\section{Diffraction of light in gravitational waves}
\begin{figure}[h]
\centering
\hspace{-0.5cm}
\includegraphics[scale=0.16]{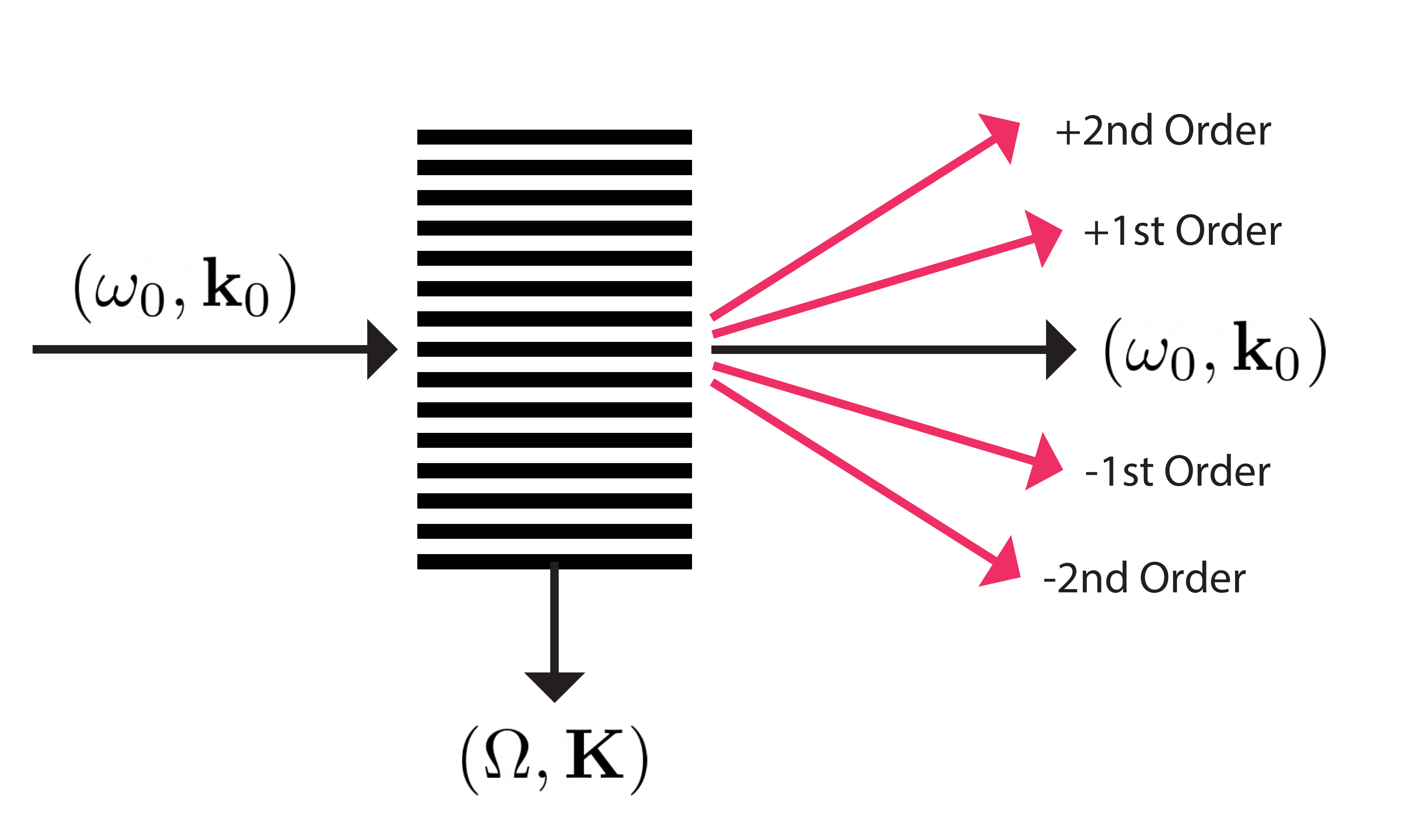}
\caption{Depiction of the interaction geometry, describing the interaction of a incident electromagnetic beam propagating in the positive x-direction with a acoustic wave propagating in the positive z-direction. Diffraction orders are colored in red.}
\label{fig:Diffraction}
\end{figure}

The phenomenon of light diffraction in an inhomogeneous and isotropic medium has been extensively studied, with foundational contributions made through the plane-wave expansion method by Raman and Nath \cite{Raman_Nath, Ghatak_Thyagarajan_1989}, along with Kogelnik's coupled-wave theory \cite{https://doi.org/10.1002/j.1538-7305.1969.tb01198.x}. A medium that is homogeneous, isotropic, and transparent to a given wavelength of electromagnetic radiation is typically characterized by a constant refractive index. However, when an acoustic wave propagates through such a medium, it induces a strain field that alters the refractive index via the photoelastic effect. This strain varies periodically with the acoustic wave, leading to a perturbative periodic modulation of the refractive index in both space and time. Such a modulation introduces spatial and temporal variations in the optical path difference of an incident plane wave, resulting in a diffraction pattern. This pattern consists of discrete diffraction orders, with the frequency of each order shifted by an integer multiple of the acoustic frequency. The case where the light and acoustic waves propagate orthogonally is illustrated in Fig. \ref{fig:Diffraction}.

In the case of gravitational waves, perturbations to the refractive index arise from the periodic variations of the spacetime metric, rather than from density fluctuations within a medium. Thus, it is postulated here that when an electromagnetic plane wave of frequency $\omega_0$ propagates through a gravitational wave of frequency $\Omega$, described by equation (\ref{DiffractionEOM}), side-band excitations $\omega_0 \pm \Omega$ are generated, each of which produce their own side band excitation in a cascading process. In this way, it is possible to consider a trial solution that is the summation of electromagnetic waves of frequency $\omega_q = \omega_0 + q\Omega$ where $q \in \mathbb{Z}$ is the diffraction order and the plane waves satisfy the dispersion relation $\omega_{q} = c k_q$.  Considering the simplest case, where we choose the incident electric field vector $\bold{E}_0$ to be orthogonal to the wave vector of the incident gravitational wave, it follows that the solution to equation.(\ref{DiffractionEOM}) with incident backward and forward propagating waves can be written as a sum of scalar waves. 

\begin{eqnarray}
 E_f(t,\bold{r}) = \sum_{q\in \mathbb{Z}} \mathcal{E}_f^{(q)}(\bold{r})e^{i(\omega_q t - \bold{\chi}_q\cdot \bold{r} )} \\ 
  E_b(t,\bold{r}) = \sum_{q\in \mathbb{Z}} \mathcal{E}_b^{(q)}(\bold{r})e^{i(\omega_q t + \bold{\chi}_q\cdot \bold{r} )}
\end{eqnarray}

where \( \mathcal{E}_f^{(q)} \) and \( \mathcal{E}_b^{(q)} \) denote the electric field amplitudes of diffraction order \( q \) generated by the forward and backward-propagating background waves, respectively, with \( \mathcal{E}_f^{(0)} = \mathcal{E}_q^{(0)} = E_0 \). The propagation vector of each order \( q \) is given by \( \boldsymbol{\chi}_q = (\boldsymbol{\alpha}_q, 0, -qK) \), where \( \boldsymbol{\alpha}_q \) corresponds to the wave number of diffraction order \( q \) along the propagation direction of the background plane waves. While it is true that plane waves are non-physical as they carry infinite energy, they can be approximated by paraxial Gaussian beams with beam widths $w_0$ and wave numbers $k_0$ within their Rayleigh ranges $Z_R = \frac{1}{2}k_0 w_0^2$ \cite{Brooker,BARNETT1994670,PhysRevA.45.8185,pampaloni2004gaussian}.

To proceed, it is necessary to explicitly define a metric for an approaching gravitational wave. Within the transverse traceless gauge (TT-gauge) the metric for a linearly polarized gravitational wave takes a simple form:

\begin{equation}
    h_{\mu \nu}^{TT} = Re[h_{+}e^{+}_{\mu \nu}e^{ik_{\mu}^g x^{\mu}}] . 
\end{equation}

 While the TT-gauge  provides an intuitive picture of the gravitational waves effects on matter and contains the radiative components of the metric perturbation, it remains the case that this is not the co-ordinate system of the observer nor the co-ordinate system in which the electromagnetic fields are expressed. Instead, the co-ordinate system of an inertial observer is described by Fermi-normal co-ordinates. The expression of the metric in Fermi-normal co-ordinates was first derived by Minser \cite{MinserFNC} and later generalized to all orders in space-time parameters by Fortini \cite{Fort}.  Given that many Beyond-Standard-Model signals are expected to reside in the sub-THz frequency regime, we consider the long-wavelength limit, where the interaction region of length \( L_p \) along the propagation direction of the incident gravitational wave satisfies \( L_p \Omega / c \ll 1 \). In this regime, the metric experienced by an inertial observer due to a gravitational wave takes the form:

\begin{gather}
    \bar{g}_{00} = 1-  R_{0k0l} x^kx^lz^n \\
    \bar{g}_{0j} = -\frac{2}{3} R_{0kjl} x^kx^lz^n \\
    \bar{g}_{ij} = \eta_{ij}-\frac{1}{3} R_{ikjl}x^kx^lz^n
\end{gather}
Noting that $k_g^{\mu}x_{\mu} = \Omega t - k_g z$ explicit expressions for the metric perturbation of the form $h_{\mu \nu} = H_{\mu \nu} (\bold{r}) cos(\Omega t - K z)$ are given:
\begin{gather}
     h_{0 0} = -\frac{ \Omega^2}{2c^2}(x^2-y^2)h_+cos(\Omega t - K z) \\
    h_{0 x} = -\frac{ \Omega^2}{3c^2}zxh_+cos(\Omega t - K z) \\
     h_{0 y} = -\frac{ \Omega^2}{3c^2}zyh_+e.cos(\Omega t - K z) 
\end{gather}

\begin{gather} 
   h_{0 z} = \frac{ \Omega^2}{6c^2}(x^2-y^2)h_+cos(\Omega t - K z) \\
    h_{xx} = -\frac{ \Omega^2}{6c^2}z^2h_+cos(\Omega t - K z)  \\
    h_{yy} = \frac{ \Omega^2}{6c^2}z^2h_+cos(\Omega t - K z) \\
     h_{zz} = -\frac{ \Omega^2}{6c^2}(x^2-y^2)h_+cos(\Omega t - K z) \\
    h_{xy} = 0 \\
    h_{xz} = \frac{ \Omega^2}{6c^2}xzh_+cos(\Omega t - K z)  \\
    h_{yz} = -\frac{ \Omega^2}{6c^2}yzh_+cos(\Omega t - K z) \label{incGWlast}.
\end{gather}
The coupled differential equations for the diffraction orders $q$ can thus be derived: 
\begin{widetext}
    \begin{eqnarray}
    \Box E^{(q)}_f(t,\bold{r}) = \frac{1}{2}\left(Q^{(q-1)}(\bold{r})e^{-i(\bold{k}_{q-1}+\bold{K})\cdot \bold{r}}+Q^{(q+1)}(\bold{r})e^{-i(\bold{k}_{q+1}-\bold{K})\cdot \bold{r}}\right) e^{i\omega_q t} \\
     \Box E^{(q)}_b(t,\bold{r}) = \frac{1}{2} \left(P^{(q-1)}(\bold{r})e^{i(\bold{k}_{q-1}-\bold{K})\cdot \bold{r}}+P^{(q+1)}(\bold{r})e^{i(\bold{k}_{q+1}+\bold{K})\cdot \bold{r}}\right) e^{i\omega_q t} 
\end{eqnarray}
\end{widetext}

Here, \( E^{(q)}_f(t,\bold{r}) \) and \( E^{(q)}_b(t,\bold{r}) \) represent the electric fields of the \( q \)-th diffraction order generated by the forward and backward propagating background electromagnetic plane waves interaction with some passing gravitational wave, respectively. The functions \( Q^{(q)}(\bold{r}) \) and \( P^{(q)}(\bold{r}) \) are determined by the metric perturbation and can be derived for each diffraction order based on the chosen interaction geometry. 

Now consider the geometry depicted in Figure (\ref{fig:Diffraction}), but with the acoustic wave replaced with a gravitational wave propagating in the z-direction and an electromagnetic beam traveling in the x-direction. Setting the initial polarization of the electric field aligned along the y-direction for simplicity, and noting that this is in the linearized regime for the metric perturbation, the equations of motion decouple and simplify significantly into a set of Helmholtz equations:

\begin{eqnarray}
    \left[  \left(\frac{\omega_{\pm}}{c}\right)^2 - \nabla^2 \right] \bold{\mathcal{E}}_f^{(\pm)}(\bold{r}) e^{-i\alpha_{\pm}x} = \frac{1}{2} Q_{\pm}^{(0)}(\bold{r}) e^{-i(\bold{k}_0 \pm \bold{K}) \cdot \bold{r}}  \\
     \left[  \left(\frac{\omega_{\pm}}{c}\right)^2 - \nabla^2 \right] \bold{\mathcal{E}}_b^{(\pm)}(\bold{r}) e^{i\alpha_{\pm}x} = \frac{1}{2}  P_{\pm}^{(0)}(\bold{r}) e^{i(\bold{k}_0 \mp \bold{K}) \cdot \bold{r}}
\end{eqnarray}

Where $q = \pm$ for the two diffraction orders generated in the linearized regime and the functions  \( Q^{(0)}(\bold{r}) \) and \( P^{(0)}(\bold{r}) \) are given by: 

\begin{eqnarray}
Q^{(0)}(\bold{r})  = \frac{E_0 h_+ k_0^2 K^2}{2} \left(  \frac{x^2-y^2}{2}+\frac{2 zx}{3} + \frac{z^2}{6}  \right)    \\ 
P^{(0)}(\bold{r})  = \frac{E_0 h_+ k_0^2 K^2}{2} \left(  \frac{x^2-y^2}{2}-\frac{2 zx}{3} + \frac{z^2}{6}  \right)  
\end{eqnarray}

Given the weak interaction between gravitational fields and light, one is able to adopt the slowly varying amplitude approximation to facilitate the analysis. It follows that:

\begin{eqnarray}
    \left[ \frac{\partial}{\partial x} - \frac{\Delta\beta_{\pm} + \nabla_{\perp}^2}{2i\alpha_{\pm}} \right] \bold{\mathcal{E}}_f^{(\pm)}(\bold{r})= \frac{iQ^{(0)}(\bold{r})}{4 \alpha_{\pm}}  e^{-i \Delta \alpha_{\pm}x \mp iKz} \\
      \left[ \frac{\partial}{\partial x} + \frac{\Delta\beta_{\pm} + \nabla_{\perp}^2}{2i\alpha_{\pm}} \right] \bold{\mathcal{E}}_b^{(\pm)}(\bold{r})= \frac{P^{(0)}(\bold{r})}{4 i \alpha_{\pm}}  e^{i \Delta \alpha_{\pm}x \mp iKz}
\end{eqnarray}

where $\Delta \alpha_{\pm} = k_0 -\alpha_{\pm}$ and $\Delta\beta_{\pm} = k^2_{\pm} -\alpha^2_{\pm}$. The solutions to the field amplitudes of the diffraction orders can be written in transverse Fourier-Space ($  FT: (y,z) \rightarrow (\eta,\xi) $) and $B(x,\boldsymbol{k}_{\perp}) = FT(\mathcal{E}(x,\boldsymbol{r}_{\perp}))$ such that:  

\begin{eqnarray}
    \bold{\mathcal{B}}_f^{(\pm)}(x,\bold{\boldsymbol{k}_{\perp}}) = \frac{i }{4 \alpha_{\pm}} \int_C e^{i\mu (\xi-x)} \Tilde{Q}_0(x,\bold{k}_{\perp}) e^{-i\Delta\alpha_{\pm} \xi} d\xi \\
    \bold{\mathcal{B}}_b^{(\pm)}(x,\bold{\boldsymbol{k}_{\perp}}) = \frac{-i }{4 \alpha_{\pm}} \int_C e^{-i\mu (\xi-x)} \Tilde{P}_0(x,\bold{k}_{\perp}) e^{i\Delta\alpha_{\pm} \xi} d\xi
\end{eqnarray}

where $C = [-L,L]$ is the domain of integration over the interaction region of length $2L$, $\mu = (\Delta \beta_{\pm} - \bold{k}_{\pm})/2\alpha_{\pm}$ and $\Tilde{Q}_0(x,\bold{k}_{\perp}) $ and $\Tilde{P}_0(x,\bold{k}_{\perp}) $ are the Fourier transform of $Q^{(0)}(\bold{r})$ and $P^{(0)}(\bold{r})$ respectively. For a laser beam whose length is very much larger than its width over the Rayleigh range, such that \( w_0 \ll L \), the dominant term is the one with the highest polynomial dependence along the propagation direction of the background beams. This leads to \( Q^{(0)}(\bold{r}), P^{(0)}(\bold{r}) \propto x^2 \). Then, using the dispersion relation $\mathcal{D}(\omega_{\pm}, \bold{k}_{\pm}) = 0$, the solutions in real space are $\bold{\mathcal{E}}_f^{(\pm)}(\bold{r}) \approx exp(\mp iK z) \Lambda(L)$ and  $\bold{\mathcal{E}}_b^{(\pm)}(\bold{r}) \approx exp(\mp i K z) \bar{\Lambda}(L)$, where the bar denotes the complex conjugate:

\begin{eqnarray}
     \Lambda_\pm(L) \approx  \int_C \xi^2 e^{-i\Delta\alpha_{\pm} \xi} d\xi 
\end{eqnarray}

It follows that solutions for the electric field amplitude of each diffraction are thus given by: 

\begin{eqnarray}
    E^{\pm}_{f}(\bold{r}) \approx \frac{i E_0 h_+ k^2_0 K^2}{16 \alpha_{\pm}} e^{i(\omega_{\pm} t -  \alpha_{\pm} x \mp K z)}\Lambda_\pm(L)  \\ 
     E^{\pm}_{b}(\bold{r}) \approx \frac{-i E_0 h_+ k^2_0 K^2}{16 \alpha_{\pm}} e^{i(\omega_{\pm} t + \alpha_{\pm} x \mp K z)}\bar{\Lambda}_\pm(L).
\end{eqnarray}

Also, $\alpha_{\pm} = \sqrt{k^2_{\pm} - K^2}$ from the dispersion relation $\omega_\pm = c k_\pm$. Thus one sees that the background forward and backward propagating electromagnetic plane wave generates two diffraction orders with frequencies $\omega \pm \Omega$ at diffraction angles $\theta_\mp$, albeit in opposite directions along the x-axis.  The diffraction itself angle is computed from the relation 

\begin{equation}
    cos(\theta_{\pm}) = \frac{\alpha_{\pm}}{k_{\pm}}.
\end{equation}

In the  low frequency  limit, the diffraction angle can be approximated by  \( \theta_\pm \approx \Omega / \omega_0 \ll 1 \).

\section{Detection scheme}

\begin{figure}[h]
\centering
\includegraphics[scale=0.28]{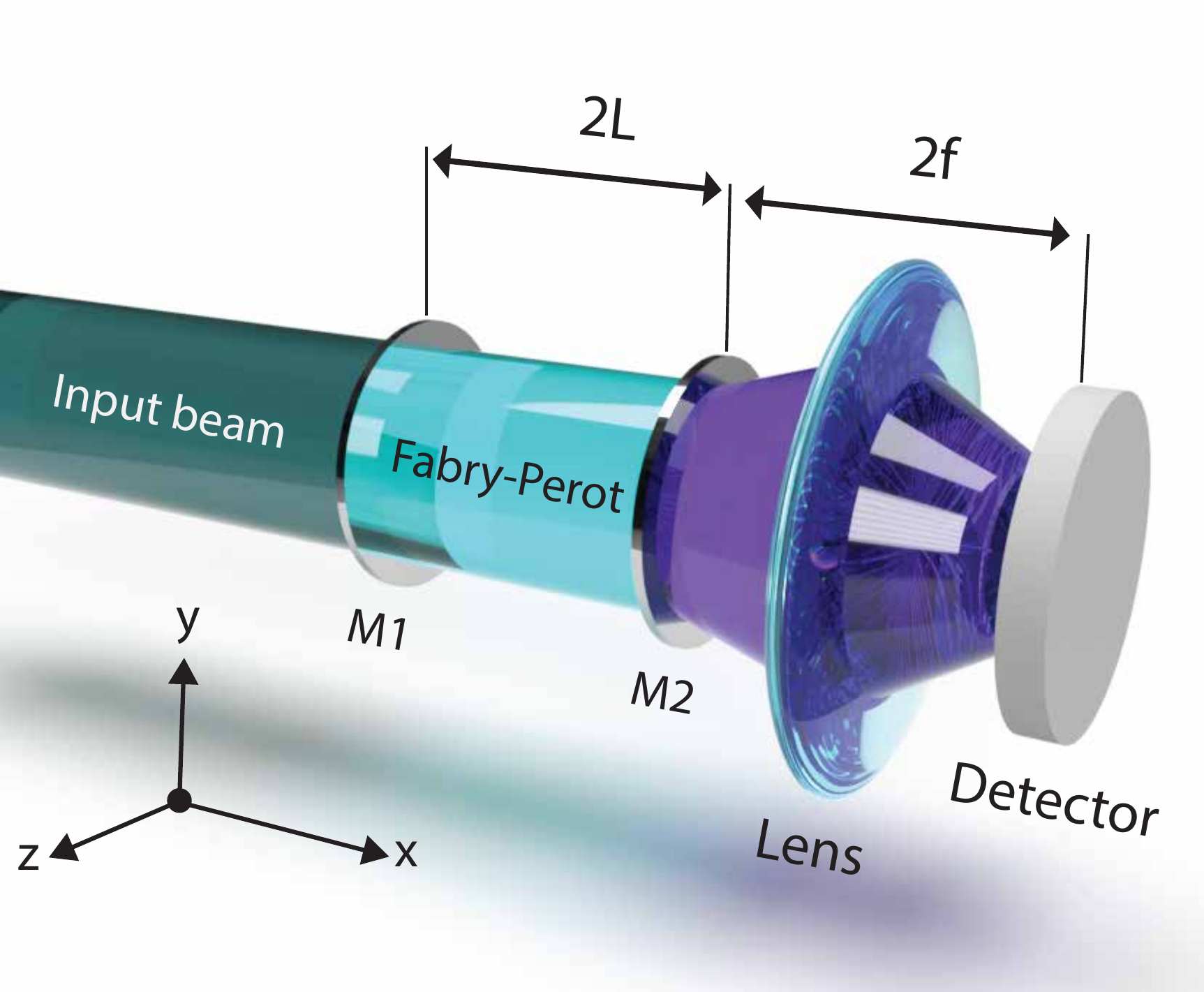}
\caption{Depiction of detection scheme geometry, describing the interaction of a incident electromagnetic beam propagating in the positive x-direction with a gravitation wave propagating in the positive z-direction. The signal  is imaged by a \textit{2f} lens geometry with $f \ll L$.}
\label{fig:tube}
\end{figure}

It is clear that the amplitude of the generated diffraction pattern and the diffraction angle are too small on their own to produce a measurable and distinguishable signal unless extremely large propagation distances from the interaction region are considered. Nevertheless, it is possible to obtain a measurable signal by first enhancing the effect through a Fabry-Pérot cavity and second detecting the imparted frequency modulation via heterodyne measurement of the beat frequency between the zeroth and first diffraction orders \cite{10.1119/1.15655,HALLAL2022100914,PhysRevD.104.L111701,narayanan2002high}. An experimental setup illustrating this concept is shown in Figure (\ref{fig:tube}).

For an electric field distribution produced at the output of a Fabry-Pérot cavity with mirrors of equal reflectivity \( R \), the field is imaged onto a detector using a \textit{2f} lens system for which $f \ll L$ such that any diffraction occurring after the Fabry-perot is negligible. The resulting field at the detector plane is given by \( U(\xi, \zeta) \), where the coordinates \( \xi \) and \( \zeta \) correspond to the detector plane:

\begin{equation}
    U_\pm(\xi, \zeta) = \frac{e^{\pm i \omega_{\pm}t}\mathcal{C}_{\lambda} }{ \lambda_{\pm} f} \int_{\mathbb{R}^2} e^{i\left( \frac{k_{\pm}}{f} [\xi z + \zeta y]\right)} e^{\mp i Kz} A(z,y) dz dy
\end{equation}

where $\mathcal{C}_{\pm} = (E_0 h_+ k^2_0 K^2 \Lambda(L)_\pm /\alpha_{\pm} \sqrt{1-R} )  $.  The factor involving the reflectivity of the mirrors arises due to the enhancement in the interactivity field of the Fabry-Pérot cavity. As the background wave propagates between the mirrors, both diffraction orders grow. Upon each reflection of the background plane wave, a new set of diffraction orders are generated, with the initial phase of the gravitational wave differing for each instance. Accounting for the initial phase of each diffraction order generated upon reflection and assuming, for simplicity, a square beam of width \( w_0 \) that satisfies the long-wavelength condition \( w_0 \Omega / c \ll 1 \) without loss of generality, the aperture function \( A(z,y) \) after \( N \) reflections is given by:

\begin{equation}
    A(z,y) = C_{N}(z) \otimes_z  \Pi(z,y)
\end{equation}

where the aperture function is the convolution of the two-dimensional (2D) top-hat function $\Pi(z,y)$ 

\begin{equation}
\Pi(z,y) =
\begin{cases} 
1, & |z| \leq \frac{w_0}{2}, \ |y| \leq \frac{w_0}{2} \\
0, & \text{otherwise}
\end{cases}
\end{equation}

and a Dirac delta comb $ C_{N}(z) $ along the z-axis: 

\begin{eqnarray}
     C_{N}(z)  = t \sum_{q=0}^{N} r^{2q}  e^{iq\Delta(\theta_{\pm})+iN\Phi_{\pm}}\Bigl(\delta(z \mp 2qd) \nonumber  \\ 
     + \delta(z \mp (2q+1)d)e^{i\frac{\Phi_{\pm}}{2}} \Bigr) 
\end{eqnarray}

and where $d = 2L sin(\theta_{\pm})$, $\Delta(\theta_{\pm}) = 4 k_{\pm} L (cos\theta_{\pm}-1) $ and $\Phi_{\pm} = 4 k_{\pm} L/c$. Defining $\Psi_\pm = \bigl[\frac{k_\pm \xi}{f}\mp K\bigr]\frac{w_0}{2} $ and $\gamma_\pm(\xi) = 2 k_\pm L \pm \Psi_\pm $ the field of imaged diffraction orders is given by:
 
\begin{eqnarray}
  U_\pm(\xi, \zeta) &=& (1-e^{i\gamma_\pm(\xi)}) \mathcal{I}_\pm(\xi, \zeta) \\
\mathcal{I}_\pm(\xi, \zeta) &=& \frac{e^{\pm i \omega_{\pm}t}\mathcal{C}_{\pm}}{ \lambda_{\pm} f} \sigma_N w^2_0 e^{i \Phi_\pm} sinc\Bigl( \Psi_\pm (\xi)\Bigr) \nonumber \\
&\times& sinc\Bigl(\frac{k_\pm w_0}{2f}\Bigr) \\ 
\sigma_N &=& \sqrt{1-R}\sum_{q=0}^N R^q e^{iq\Delta(\theta_{\pm})}e^{\pm 2qi\Psi_\pm}
\end{eqnarray}
 
It follows that the combined field of both enhanced $\pm$ diffraction orders on the detector on the detector is given by $U(\xi, \zeta) = U_+(\xi, \zeta)+U_-(\xi, \zeta)$. For observation times exceeding the cavity build-up time $\tau_{build} = L/(1-R)c $ the time-averaged intensity then can be approximated by $I(\xi, \zeta) = \lim_{N\to\infty} \bigl<|U(\xi, \zeta)|^2 \bigr>_T \epsilon_0 c$ such that: 

\begin{eqnarray}
      I(\xi, \zeta)  = \sum_{j \in \pm} \frac{2 \epsilon_0 c |\mathcal{C}_{j}|^2 w^4_0}{(1-R)\lambda^2_j f^2} sinc^2(\Psi_j(\xi))  \nonumber \\
      \times \frac{sinc^2(\frac{k_j w_0}{2f}\zeta) sin^2(\gamma_j(\xi)/2)}{1+ \mathcal{F} sinc^2(\psi_j(\xi))}.
\end{eqnarray}

Here, $\mathcal{F} = 4R/(1-R)$ is the finesse of the cavity,  $\gamma_\pm(\tau) = 2 k_\pm L \pm 2 \tau \frac{L}{w_0}sin(\theta_{\pm}) $  and $\psi_\pm(\tau) = \Delta(\theta_{\pm}) 
 \pm 4 \tau \Psi_\pm d/w_0$. Thus, by integration over $\mathbb{R}^2$ in the plane of the detector, the time-averaged power\textit{ P} of the signal is given by: 

\begin{equation}
    P = \sum_{j \in \pm}  \frac{2 P_0 k^4_0 K^4 h^2_+}{(1-R)^2 \alpha_\pm^2 } \Lambda^2_j(L)  \mathcal{G}_j(w_0, L) 
\end{equation}

where $P_0$ is the power of the background plane wave. 

$\Lambda_\pm(L)$ is evaluated and the function $ \mathcal{G}_j(w_0, L) $ is given.

\begin{eqnarray}
     \Lambda_\pm(L) =    \left( \frac{L^2}{\Delta \alpha_\pm} - \frac{2}{\Delta \alpha^3_\pm}\right) sin(\Delta \alpha_\pm L) \nonumber \\ 
    + \frac{2L}{\Delta \alpha^2_\pm} cos(\Delta \alpha_\pm L)  \\ 
   \mathcal{G}_\pm(w_0, L) = \frac{1}{\pi} \int_{\mathbb{R}} \frac{Sinc^2(\tau)Sin^2(\gamma_\pm(\tau)/2)}{1+\mathcal{F} sin^2(\psi_\pm(\tau))} d\tau
\end{eqnarray}

Here, the substitution \( \tau = \bigl[\frac{k_\pm \xi}{f} \mp K\bigr] \) has been used for which $\gamma_\pm(\tau) = 2 k_\pm L \pm 4 \tau \frac{L}{w_0}sin(\theta_{\pm}) $  and $\psi_\pm(\tau) = 4 k_\pm L(cos(\theta_\pm)-1) \pm 8 \tau \frac{L}{w_0}sin(\theta_{\pm})$. 

For heterodyne measurements, in the scenario where shot noise is the dominant noise contribution, the signal to noise ratio reduces to that of directly measuring the shot-noise-limited signal $S/N = \sqrt{ P T/ \hbar \omega_0}$, where $T$ is the observation time \cite{10.1119/1.15655,HALLAL2022100914,PhysRevD.104.L111701,narayanan2002high}. However, a key quantity in characterizing detector performance is the spectral noise density \( S_n(\Omega) \), which defines the noise floor and enables estimation of sensitivity to incoherent signals. It can be determined by expressing the signal to noise ratio in terms of \( S_n(\Omega) \) as \( S/N = \sqrt{h_+^2 T / S_n(\Omega} \) \cite{Maggiore:2007ulw,Moore_2015}:

\begin{eqnarray}
    S^{\frac{1}{2}}_\Omega = \frac{(1-R)}{k^2_0K^2}\sqrt{\frac{\hbar}{2 P_0} \sum_{j\in \pm} \frac{\omega_j}{\Lambda^2_j(L) \mathcal{G}_j(w_0, L) }}.
\end{eqnarray}

In order to determine the maximum enhancement capabilities of using a Fabry-Pérot cavity and to ensure that the signal generated by the proposed gravito-optic Hetrodyne detector is composed entirely of the Heterodyne signal rather than fluctuations in the laser frequency or arm length due to vibration from the environment, currently and future experimental and technological constraints have to be taken into account.  

Conventional laser mirrors typically have reflection coefficients of 0.99, corresponding to a cavity finesse of a few hundred. However, optical super-mirrors, typically dielectric in nature, can be optimized for exceptionally high reflectance. Fabry-Pérot cavities constructed with such mirrors can achieve finesse values as high as $\mathcal{F} = \mathcal{O}(10^6)$ in the mid-infrared to optical frequency spectrum, enabling highly sensitive detection capabilities \cite{sones1997optimization,Truong,Muller:10,Meng:05}. Advanced LIGO, for is capable of achieving a fractional arm-length change due to the passage of a gravitational wave on the order of  \(\Delta L / L < 10^{-23}\) and maintains fractional frequency deviations below 1 Hz \cite{Izumi:14}. Cosmic Explorer is designed to meet or surpass these benchmarks, featuring arm lengths exceeding \(40 \, \text{km}\) and circulating power of \(125 \, \text{MW}\) within its Fabry-Pérot cavities \cite{hall2022cosmic}. The laser interferometer space antenna takes this even further and considers arm-lengths of 2.5 million kilometers. Additionally, such precision is maintained over entire observing runs, which can persist for several months to a year far exceeding the estimated observation time required for a detection in the proposed scheme. Based on the currently functional and proposed technologies, a set of viable experimental parameters for the evaluation of the performance of the proposed detection scheme have been outlined in table.\ref{Tab:Laserparameters}.

\begin{table}[H]
\begin{center}
   \begin{tabular}{|c|c|c|c|}
\hline
Name  & 2L    & $\mathcal{F}$   & $P_0$ \\ \hline
A     & 200m  & $10^6$          & 1W    \\ \hline
B     & 2km   & $10^5$          & 1W    \\ \hline
C     & 20km  & $10^4$          & 1W    \\ \hline
LIGO  & 3km   & $5 \times 10^3$ & 150W  \\ \hline
\end{tabular} 
\end{center}
\caption{System parameters.}
\label{Tab:Laserparameters}
\end{table}

\begin{figure}[h]
\centering
\hspace{-0.5cm}
\includegraphics[scale=0.65]{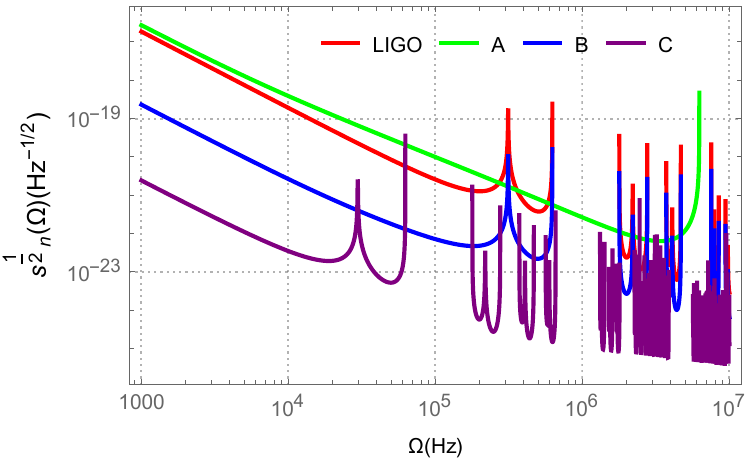}
\caption{Spectral noise density curves for hypothetical ground based detectors ranging from 200m to 20km. A curve for a detector based on LIGO's parameters has been included for reference.}
\label{fig:GroundSND}
\end{figure}

Considering a laser operating at a wave-length of $1053 \text{nm}$ the frequency dependence of the spectral noise density for the different specifications outlined in table.\ref{Tab:Laserparameters} is given in figure.{\ref{fig:GroundSND}}. 

It is evident that increasing the length and finesse of the detector enhances its sensitivity, albeit at the expense of bandwidth. The rapid oscillatory behavior at higher gravitational wave frequencies arises from phase velocity mismatch, encoded in the functions \( \Lambda_j(L) \), and the interference of the Fabry-Pérot-enhanced signal, captured by \( \mathcal{G}_j(L, w_0) \). For clarity, the frequency domain is restricted to 10 MHz. However, the ultimate frequency limit of heterodyne detection is set by the electronic response time of the data acquisition system which is 10 GHz.

\section{Transient gravitational wave detection}

Compact binary systems serve as sources of transient gravitational wave radiation, with energy loss due to gravitational wave emission driving the inspiral of the constituent objects and ultimately leading to their coalescence. The resulting waveform, characterized by a chirping frequency evolution, encodes essential information about the system’s masses, spins, and orbital dynamics. While black hole and neutron star mergers constitute the primary astrophysical sources of gravitational waves, a variety of beyond-Standard-Model scenarios predict the existence of exotic compact objects as potential sources \cite{Cardoso,Giudice_2016,PhysRevD.110.024034,Raidal:2017mfl,Franciolini:2022htd,Ireland:2023avg,Dong:2015yjs}. These hypothetical objects, arising in diverse new-physics frameworks, provide compelling targets for gravitational wave searches and offer novel avenues for probing fundamental physics.

In the final stages of the inspiral, a compact binary system radiates a few percent of its total mass in gravitational waves, corresponding to an immense energy release. Despite the intrinsic strength of the emitted signal, it is expected to be embedded within a significantly larger noise background. However, the prolonged duration of the inspiral phase allows for coherent signal tracking over numerous cycles in a broadband detector, enabling the signal-to-noise ratio (SNR) to be accumulated to detectable levels. The matched-filtered SNR is given by \cite{Maggiore:2007ulw,Moore_2015, Aggarwal:2020olq}:

\begin{equation}
\text{SNR}^2 = 4 \int_{0}^{\infty} \frac{|\tilde{h}(\Omega)|^2}{S_n(\Omega)} \, d\Omega
\end{equation}

Where \(\tilde{h}(\Omega)\) denotes the spectral amplitude of the gravitational wave signal. During the inspiral phase of a merger between two compact objects of equal mass, the spectral amplitude of the gravitational wave signal is given by \cite{Maggiore:2007ulw,Aggarwal:2020olq}:

\begin{equation}
h_c(\Omega) = \frac{c}{D} \sqrt{\frac{5}{24} } \pi^{-2/3} \left(\frac{G \mathcal{M}_c}{c^3}\right)^{5/6}\, \Omega^{-1/6}
\end{equation}

Where \( D \) and \( \mathcal{M}_c \) denote the binary system's distance from the detector and chirp mass, respectively. For a binary system consisting of equal-mass compact objects, the chirp mass is given by:

\begin{equation}
    \mathcal{M}_c =  2^{-1/5} M_{BH}
\end{equation}

The given spectral amplitude remains valid until the compact objects in the binary system reach the innermost stable circular orbit (ISCO), beyond which the inspiral phase terminates, with the emitted gravitational waves at this point having a corresponding frequency \(\Omega_{ISCO}\) \cite{Maggiore:2007ulw,Aggarwal:2020olq}:

\begin{equation}
\Omega_{ISCO} = 4.4 \pi \left( \frac{M_{\odot}}{M_{BH}}\right)
\end{equation}

\begin{figure}[h]
\centering
\hspace{-0.5cm}
\includegraphics[scale=0.54]{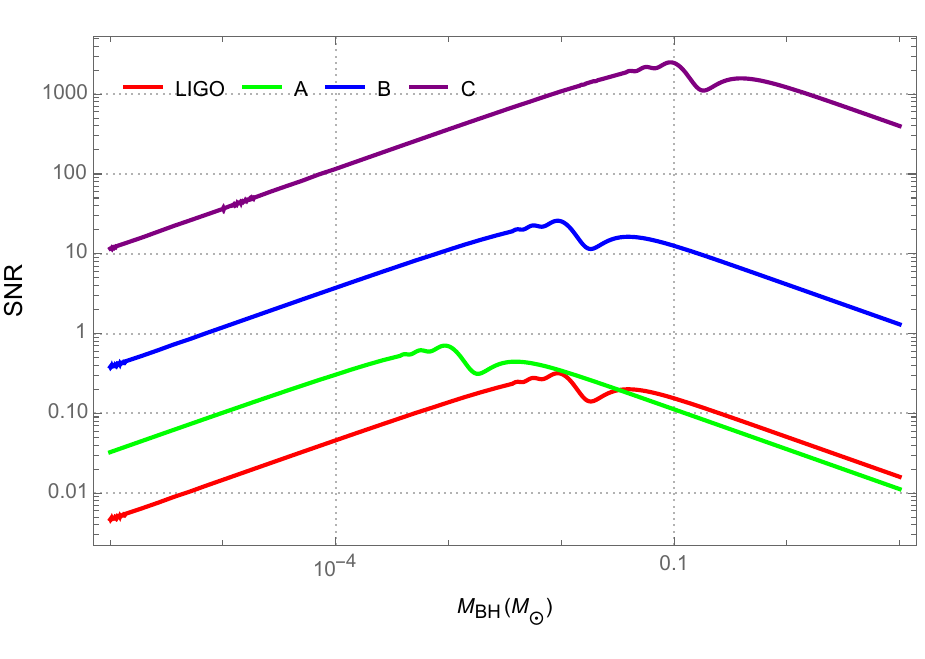}
\caption{Signal to noise ratio curves for transient signals in hypothetical ground-based detectors with of lengths ranging from 200 m to 20 km (see Table \ref{Tab:Laserparameters}). For reference, an additional curve corresponds to a detector with LIGO-like parameters was added. All curves are computed for a distance of \( D = 30 \) kpc and wavelength \( \lambda = 1053 \) nm. The analysis considers source masses in the range \( 10^{-6} M_{\odot} \)–\( 10 M_{\odot} \).}
\label{fig:GWsignal}
\end{figure}

Using the previously derived equation for the spectral noise density and limiting the integration domain to the gravitational wave frequency at the ISCO, the estimated detector performance, parameterized by the signal-to-noise ratio as a function of gravitational wave frequency for compact binary systems at a distance of \(30 \text{kpc}\), is presented in Fig. (\ref{fig:GWsignal}).

The primary astrophysical mechanism for black hole formation is the collapse of stars. Whether a star collapses into a black hole depends on its mass relative to the Tolman–Oppenheimer–Volkoff limit. Early estimates of this limit ranged between \(2.2\) and \(2.9 \, M_\odot\) \cite{bombaci1996maximum, Kalogera_1996}. However, further refinements, including the first gravitational wave detection from merging neutron stars, GW170817, and considerations of the progenitor star's rotation, have led to an updated lower limit for the mass of the resulting black hole, estimated to be \(2.6 \, M_\odot\) \cite{Pooley_2018, doi:10.1126/science.359.6377.724, Rezzolla_2018}. Therefore, the detection of compact binaries in the sub-solar mass range that are not of neutron star origin, confirmed through a multi-messenger approach, could indicate the presence of exotic compact objects \cite{Cardoso,Giudice_2016,PhysRevD.110.024034} or primordial black holes \cite{Raidal:2017mfl,Franciolini:2022htd,Ireland:2023avg,Dong:2015yjs}.

\section{Discussion}

The lowest-order correction to the electromagnetic wave equation in the low frequency  limit due to a perturbed space-time has been considered. Within the wave optics formalism, it was shown that in the presence of a gravitational wave, electromagnetic radiation undergoes a diffraction process analogous to acousto-optic diffraction, leading us to designate this effect as *gravito-optic diffraction*. In the specific case where the gravitational wave and electromagnetic wave propagate in orthogonal directions, it was found that, in the low frequency limit, the diffraction angle of the gravito-optic effect is exactly the same as that produced if the gravitational wave were replaced by an acoustic wave. The gravito-optic effect itself is too weak to be directly detected; however, a detection scheme based on this effect, enhanced by a Fabry–Pérot cavity and measured via heterodyning of the enhaced signal with the zeroth diffraction order referred to as the gravito-optic heterodyne detector was considered. The sensitivity of this detector increases with both the length and finesse of the cavity at constant power, albeit at the expense of detector bandwidth, as evident from the spectral noise density plot. Despite this trade-off, when analyzing transient gravitational wave signals from coalescing compact binaries, an increase in the detector length extends the range of sub-solar mass compact objects that can be probed. Consequently, this detection approach may provide a novel strategy for detecting or constraining the abundance of sub-solar mass compact objects.

\section{Summary and Future Work}

The required length to achieve a SNR exceeding unity is on the order of kilometers or greater. In a companion paper "Search for black hole super-radiance using gravito-optic hetrodyne detection" \cite{Atonga_GravitoOptic}, it was also demonstrated that the proposed detection architecture necessitates the use of a kilometer-scale detector to probe gravitational waves generated by boson annihilation around black holes \cite{PhysRevD.81.123530,PhysRevD.83.044026}. Thus, given that the technology necessary for constructing a gravito-optic heterodyne detector closely resembles that of existing and proposed gravitational wave detectors, an intriguing question arises: could current and future observatories, such as LIGO or Cosmic Explorer, both designed for the sub-10 kHz frequency band, be adapted to also probe the high-frequency gravitational wave spectrum, up to 10 GHz? In this present article, only the case of orthogonal propagation has been considered and the results are indeed very promising. The determination of the effective detection volume needed to estimate the expected number of detection events for different compact objects requires a fuller understanding of the angular dependence of the signal-to-noise ratio. This aspect will be the focus of future work.

\section{Acknowledgements} 

The authors gratefully acknowledge useful discussions with Dr. Aur\'elien Barrau and Dr. Killian Martineau from Université Grenoble Alpes and members of the Norreys' research group in the Clarendon Laboratory. They also thank all of the staff of the Central Laser Facility, Rutherford Appleton Laboratory for their assistance in the development of this work. This research was funded in whole or in part by the Oxford-ShanghaiTech collaboration agreement and UKRI-STFC grant ST/V001655/1. For the purpose of Open Access, the authors have applied a CC BY public copyright licence to any Author Accepted Manuscript (AAM) version arising from this submission.

\bibliography{refs.bib}

\end{document}